\documentclass[a4paper,fleqn,usenatbib]{mnras}

\usepackage{newtxtext,newtxmath}
\usepackage[T1]{fontenc}
\usepackage{ae,aecompl}

\usepackage{graphicx}	% Including figure files
\usepackage{amsmath}	% Advanced maths commands
\usepackage{amssymb}	% Extra maths symbols
\usepackage{pdflscape}
\usepackage{ulem}
\usepackage{array}
\usepackage[dvipsnames]{xcolor}
\newcommand{\lta}{\mathrel{\hbox{\raise 0.4 ex \hbox{$<$}\kern
                   -1.5 ex\lower .4 ex\hbox{$\sim$}}}}
\newcommand{\gta}{\mathrel{\hbox{\raise 0.4 ex \hbox{$>$}\kern
                   -1.5 ex\lower .4 ex\hbox{$\sim$}}}}

\title[The SVP method for radiative accelerations]{An improved parametric method for evaluating radiative accelerations in stellar interiors}

\author[G. Alecian \& F. LeBlanc]{
G.~Alecian$^{1}$\thanks{E-mail:georges.alecian@obspm.fr},
F.~LeBlanc$^{2}$
\\
$^{1}$LUTH, CNRS, Observatoire de Paris, PSL University, Universit{\'e} Paris Diderot, 5 Place Jules Janssen, F-92190 Meudon, France\\
$^{2}$D{\'e}partement de physique et d'astronomie, Universit{\'e} de Moncton, Moncton, NB, E1A 3E9, Canada\\
}

\date{Accepted 2020 August 19. Received 2020 August 18; in original form 2020 July 21.
}

\pubyear{2020}

\begin{document}
\label{firstpage}
\pagerange{\pageref{firstpage}--\pageref{lastpage}}
\maketitle

\begin{abstract}
The single-valued parameter (SVP) method is a parametric method that offers the possibility of computing radiative accelerations in stellar interiors much faster than other methods. It has been implemented in a few stellar evolution numerical codes for about a decade. In the present paper, we describe improvements we have recently brought in the process of preparing, from atomic/opacity databases, the SVP tables that are needed to use the method, and their extension to a larger stellar mass domain (from 1 to 10 solar mass) on the main-sequence. We discuss the validity domain of the method. We also present the website from where new tables and codes can be freely accessed and implemented in stellar evolution codes.
\end{abstract}


%%%%%%%%%%%%%%%%%%%%%%%%%%%%%%%%%%%%%%%%%%%%%%%%%%

%%%%%%%%%%%%%%%%% BODY OF PAPER %%%%%%%%%%%%%%%%%%

\section{Introduction}
\label{sec:intro}

It is well established since \citet{MichaudMi1970y} that abundance anomalies in upper main-sequence chemically peculiar Population I stars are caused mainly by atomic diffusion. It is also presently well known that atomic diffusion can affect elements distribution in any type of stars provided that large scale motions (such as convection) that mix the medium are weak enough \citep[see a complete discussion in][]{MichaudMiAlRi2015}.

Shortly after the pioneering work of \citet{MichaudMi1970y} that addresses the case of abundance anomalies in Ap stars atmospheres, atomic diffusion and radiative accelerations were essentially considered in atmospheres of chemically peculiar stars \citep[for instance on the 1970's'][]{MichaudMiReCh1974,AlecianAl1977,VauclairVaHaPe1979,BorsenbergerBoMiPr1979}. The most recent works in this field have led to atmospheric modelling of blue horizontal-branch stars \citep{Hui-Bon-HoaHuLeHa2000c,LeBlancetal2010}, as well as magnetic ApBp and HgMn stars, for which time-dependent calculations of diffusion for modelling 3D distribution of metals in magnetic CP star atmospheres \citep{AlecianAlSt2017}, and also calculations that include mass loss has been achieved to model the abundance stratification build-up \citep{AlecianStift2019}. Such modelling aims to better understand observational anomalies (abundances, superficial distribution of the elements, photometric jumps and gaps, etc.) detected for various types of chemically peculiar (CP) stars. For instance, photometric jumps \citep{Grundahletal1999} and gaps  \citep{Ferraroetal1998} have been detected for hot horizontal-branch stars. These anomalies have been shown to be caused by atmospheric structure changes due to the vertical stratification of metals created by atomic diffusion (\citealt{Hui-Bon-HoaHuLeHa2000c}; \citealt{LeBlancetal2009}; \citealt{LeBlancetal2010}). Since the framework of the study discussed in this paper concerns the interior of stars, we will not consider stellar atmospheric modelling hereafter.

Particle transport can affect the structure of stars (mainly through opacity effects), and so, it may also modify their evolution. Atomic diffusion has been included in several stellar evolution codes to study its effect on the evolution of several types of stars, on their pulsation patterns, and to predict their surface abundances. Examples of such calculations are given below.

In most cases, atomic diffusion is principally regulated by the competition between gravity and radiative accelerations (${g}_{\rm rad}$). Radiative accelerations of the various species present in stars are due to momentum transfer from photons to atoms following bound-bound or bound-free transitions. Their computation can be quite heavy since they depend on the total absorption cross section of the species and on the local physical conditions. An integration of the product of the photons absorption cross section times the radiative flux over frequency is needed (for instance, see Eq.~1 of \citealt{AlecianAlLe2000}). Radiation flux at a given frequency depends on the total monochromatic opacity, therefore the ${g}_{\rm rad}$ of a given species depends on this  total opacity, in addition to its specific contribution to the opacity. Extensive atomic data bases are therefore required for a more precise evaluation of ${g}_{\rm rad}$.

At large optical depths, the calculation of the radiation flux is much simpler than in optically thin media (atmospheres), since the {\it diffusion approximation} \citep{Milne27} can apply, and then the photon flux can be estimated for local physical conditions. However, despite this simplification, computation of radiative accelerations in stellar interiors were quite difficult to carry out before the availability (i.e. before the last decade of the previous century) of large atomic and opacity databases, because there was a huge lack of atomic data for highly ionised atoms. First estimates of radiative accelerations in stellar interiors after \citet{MichaudMi1970y} were obtained by \citet{MichaudMiChVaetal1976} who had developed approximate formulae for radiative accelerations. To overcome the lack of atomic data, a first version of a parametric method for radiative accelerations was developed by \citet{AlecianAl1985} and \citet{AlecianAlAr1990a} and this method allowed, when necessary, to extrapolate the ${g}_{\rm rad}$ of low charged ions (with known atomic data) to highly ionised ions along the isoelectronic sequences. This parametric method was also used by \citet{AlecianAlMiTu1993} to calculate the ${g}_{\rm rad}$ of Fe using atomic data from The Opacity Project (OP) data prior their public release \citep{SeatonSeZeTuetal1992}. This work was the precursor of the use of large atomic databases for the calculation of radiative accelerations.

\citet{GonzalezGoLeAretal1995} developed a mixed method in which direct integration over each line profile is done where the contribution of the transition under consideration is subtracted from the total opacity, therefore giving an approximation of the background opacity (through low resolution sampling). The background opacity being the opacity due to all other sources except for the transition under consideration. In their procedure, the total opacity used to estimate the background opacity is approximated by distributing the opacities to 4000 evenly spaced intervals for 0 $< u <$ 20 where $ u = h\nu/kT $. These values of total opacities are pretabulated on  two-dimensional grids of temperature and $R_e = N_e/T^3$ (where $T$ is the temperature and $N_e$ the electron density). This method was used to evaluate the ${g}_{\rm rad}$ of CNO \citep{GonzalezGoLeAretal1995} and Fe \citep{LeBlancLeMi1995i}, and opened the way of the sampling methods.

Another method for calculating ${g}_{\rm rad}$ is by properly sampling the radiative flux on a sufficiently fine frequency grid so that the value obtained after integration over the radiation frequency converges to a value that is stable relative to an increase of the grid resolution. In this procedure, the whole opacity spectrum of a given species may be treated simultaneously instead of treating each transition separately. This method is commonly called the opacity sampling method. The chosen frequency grid must be fine enough to properly sample the atomic lines (i.e. \citealt{LeBlancLeMiRi2000}) which are as a whole the dominant contribution to ${g}_{\rm rad}$ as compared to bound-free transitions. The Montreal evolution code, which incorporates atomic diffusion (\citealt{TurcotteTuRiMi1998r}; \citealt{RichardRiMiRi2001l}; \citealt{VickViMiRietal2010}) employs the opacity sampling method. Monochromatic opacities of OPAL \citep{IglesiasIgRo1996} which are calculated on a 10000-point frequency grid are used for their modelling. 

\citet{SeatonSe1997,SeatonSe2005,SeatonSe2007u} published a method for evaluating ${g}_{\rm rad}$ in stellar interiors that consists of pretabulated values on two-dimensional grids of temperature and the density of free electrons. These interpolation tables are then used to obtain the ${g}_{\rm rad}$ of a given species at a given point in a stellar model. The atomic data used are those from The Opacity Project \citep{SeatonSeZeTuetal1992} and the tables and related computer routines are available from TOPbase \citep{CuntoCuMeOcetal1993}. The underlying method used to calculate these pretabulated values for ${g}_{\rm rad}$ is the opacity sampling procedure (with a frequency resolution of $10^{5}$ points). Since the integration on frequency is already done when using these two-dimensional grids, it is less numerically onerous than direct opacity sampling.

Another method for calculating ${g}_{\rm rad}$ that has the quality of being very numerically efficient is the SVP (standing for Single-Valued Parameters) method (\citealt{AlecianAlLe2002}; \citealt{LeBlancLeAl2004}). This method is based on the parametric equations developed by \cite{AlecianAl1985}, \cite{AlecianAlAr1990a}  and \cite{AlecianAl1994}. 

The SVP method uses parametric equations (see \citealt{AlecianAlLe2002}) that separate the terms depending explicitly on atomic data from those depending on the abundance of the species under consideration. The parameters found in theses equations are pretabulated for each species treated and are calculated using OP atomic data \citep{LeBlancLeAl2004}. \citet{AlecianAlLeMa2013} also calculated the ${g}_{\rm rad}$ of Sc with this procedure while using the atomic data calculated by \citet{MassacrierMaAr2012}. The use of these parametric equations drastically diminish the computing time needed for ${g}_{\rm rad}$ calculations (about $10^3$ times faster than the OPCD codes) and are therefore well suited for evolutionary models. The price of the higher computing performance is the lost of the accuracy that is estimated to be better than 0.3 dex in average for $\log{{g}_{\rm rad}}$ when the abundance is within $\pm{1.0}$ dex the solar value, and much more accurate for a solar abundance. We have adopted this value of 0.3 dex as an acceptable error on radiative accelerations, that allows sufficiently accurate modelling of stellar evolution. We are reassured about this choice, by several comparisons of numerical codes using various method in computing radiative accelerations (see below). More details concerning these parametric equations are given in Sec.~\ref{sec:svp}. 

The SVP method\footnote{The method has been presently used only for main-sequence stars. For stars outside de main-sequence, specific tables will certainly be required (see Sec.~\ref{sec:comments}).}, for calculating ${g}_{\rm rad}$ has been implemented in the Toulouse-Gen\`{e}ve Evolution Code or TGEC \citep{Hui-Bon-HoaHu2008j} and has been applied to study thermohaline convection \citep{TheadoThVaAletal2009} in A-type stars. A more updated version of the TGEC code with atomic diffusion is described in \citet{TheadoThAlLeetal2012h} and it has been recently applied to study the consequences of atomic diffusion in A-type stars \citep{DealDeRiVa2016}. Meanwhile, \cite{DealDeAlLeetal2018} have implemented atomic diffusion in the CESTAM \citep{MarquesMaGoLeetal2013r} evolution code using the SVP method to study solar-like oscillations in main-sequence stars. More recently, \citet{DealDeGoMaetal2020k} have used this code to study the combined effects of rotation and atomic diffusion on chemical mixing in low-mass stars. Several test cases carried out by these various codes \citep[see for instance][]{DealDeAlLeetal2018} have shown that results are compatible with those obtained by the Montreal code \citep{TurcotteTuRiMietal1998}.

The present paper describes improvements brought to the SVP method. The next section (Sec.~\ref{sec:svp}) will briefly recall the basic equations defining the method in question. This will be followed (Sec.~\ref{sec:improv}) by the description of the many improvements proposed. Sample results will then be presented as well as data and codes that are publicly available on the internet(Sec.~\ref{sec:sampl} and Sec.~\ref{sec:data}). 
Section ~\ref{sec:comments} presents some additional comments about the present work, and on the SVP method (including details on its validity domain).

\section{The single-valued parameter (SVP) method}
\label{sec:svp}
As mentioned above, the main goal for developing the SVP method is to have a procedure to calculate ${g}_{\rm rad}$ that is numerically efficient and relatively easy to implement in existing stellar evolution codes. In the SVP formulae, the terms explicitly dependent on the atomic data are separated from those depending on the abundance of the species under consideration. This method consists in having only 6 parameters per ion for a given stellar mass, which enter a few simple formulae that are functions of the local concentration of the element and that give a good approximation of radiative accelerations that is sufficiently precise for most applications in stellar modelling. Therefore, small tables are sufficient to calculate ${g}_{\rm rad}$ and once the six parameters are obtained and made available to end users for each ion (see Sec.~\ref{sec:data}), ${g}_{\rm rad}$ calculations are numerically economical, since complete atomic or opacity data are no longer required. In this section, the main equations related to this method are summarised for both bound-bound and bound-free transitions. The detailed development can be found in \citet{AlecianAlLe2002} and \citet{LeBlancLeAl2004}.

\subsection{Bound-bound transitions}
\label{sec:bb}

Several assumptions are made to lead to simplified parametric equations. First, as for the other methods mentioned in Sec.~\ref{sec:intro}, these equations are developed for large optical depths (i.e. deeper than the atmosphere) and therefore the {\it diffusion approximation} \citep{Milne27} can apply. This simplifies the treatment of the monochromatic radiative flux.  Also, for the acceleration due to bound-bound transitions, it is assumed in a first step that all atomic lines saturate like Lorentz line profiles, which allows an analytical integration over the frequency of the integral found in the expression of  ${g}_{\rm rad}$. The contribution of the transition under consideration to the monochromatic total opacity is identified and its effect is estimated through a first order Taylor expansion. This finally leads to the simplified formulae. \citet{AlecianAlLe2002} and \citet{LeBlancLeAl2004}, based on the previous work of \citet{AlecianAl1985} and \citet{AlecianAlAr1990a}, found that the ${g}_{\rm rad}$ due to bound-bound transitions may be expressed as:

\begin{equation}
g_{\rm{i,line}}=q\varphi_{i}^{*}\left({1+\xi_{i}^{*}C_{i}}\right){\left({1+{{C_{i}}
\over{p\psi_{i}^{*2}}}}\right)}^{\alpha_{i}}
\label{gline}
\end{equation}

\noindent where
\begin{equation}
q=5.575\times 10^{-5}{{T_{\rm
eff}^{4}}\over{T}}{\left({{{R}\over{r}}}\right)}^{2}{{1}\over{A}}
\label{q}
\end{equation}

\noindent and
\begin{equation}
p=9.83\times 10^{-23}{{N_{e}T^{-{{1}/{2}}}}\over{X_{H}}}.
\label{b}
\end{equation}

\noindent The quantities $T_{\rm eff}$ and $R$ are the effective temperature and
radius of the star, while $T$ and $r$ are the local temperature and radius. The variable $N_e$ represents the
density of free electrons,  $X_H$ the hydrogen mass fraction, $A$ the atomic mass in
atomic units of the species under consideration and $C_i$ the concentration (in
number) of the ion relative to hydrogen. It should be noted that the expression found in Eq.~\ref{b} is now defined by the variable $p$ while it was $b$ in previous publications (i.e. \citealt{AlecianAlLe2002}). This modification is made to avoid confusion with the variable $b{_{i}}$ found below in Eq.~\ref{gcontchi}. The parameters $\varphi{_{i}}^{*}$,
$\psi{_{i}}^{*}$ and $\xi{_{i}}^{*}$ are the values of $\varphi{_{i}}$ ,
$\psi{_{i}}$ and $\xi{_{i}}$ (see below)
calculated at the stellar model layer where the relative population of ion $i$ is near its maximum (see equation 8 of \citealt{LeBlancLeAl2004}). We recall the definition of $\varphi{_{i}}$ ,
$\psi{_{i}}$ and $\xi{_{i}}$ \citep{AlecianAlLe2002}:

\begin{equation}
\varphi{_{i}}={g_{i,0}\over q},
\label{phi}
\end{equation}

\noindent
where $g_{i,0}$ is the radiative acceleration of ion $i$ when its concentration is vanishing (Eq.~11 of \citealt{AlecianAlLe2002}), and:

\begin{equation}
\psi{_{i}}=\left({C_{i,S}\over p}\right)^{{1}/{2}}
\label{psi},
\end{equation}

\noindent
where $C_{i,S}$ is the concentration of ion $i$, above which saturation of lines is strong (Eq.~14 of \citealt{AlecianAlLe2002}).

Parameter $\xi{_{i}}^{*}$ is defined by Eq.~15 of \citealt{AlecianAlLe2002}.

The parameters $\varphi{_{i}}^{*}$, $\psi{_{i}}^{*}$ and $\xi{_{i}}^{*}$ are therefore calculated at a single value of temperature and density for a given stellar mass (which is the source of the acronym SVP for the method discussed here). The parameter $\varphi{_{i}}^{*}$ depends on the oscillator strengths of the bound-bound transitions of the ion. The parameter $\psi{_{i}}^{*}$ is related to the line widths and therefore accounts for the saturation effect. The third parameter $\xi{_{i}}^{*}$ depends on the contribution of the ion to the total opacity and also affects to a lesser extent (see Sec.~\ref{sec:unfit}) the dependence of the acceleration on abundance. These three parameters are calculated with the atomic data from The Opacity Project.

As we previously mentioned, the parametric equation (Eq.~\ref{gline}) was developed for pure Lorentzian profiles and because, in this assumption, all line widths saturate according to the inverse of the square root of the abundance, the parameter $\alpha{_{i}}$ in Eq.~\ref{gline} was first set equal to -0.5 (it would have been close to -1.0 for pure Gauss profile) by \citet{AlecianAl1985}. Actually, in the SVP method, this parameter $\alpha{_{i}}$ is allowed to be adjusted to take into account the fact that the lines are not all Lorentzian in nature. Therefore, to mimic the effect of mixture of Voigt profiles (since many lines of various strength are involved), we adjust the value of $\alpha{_{i}}$ by fitting the SVP accelerations to those calculated by a more precise method (see Sec.~\ref{sec:fit}). In our method, we presently use the accelerations of \cite{SeatonSe2007u} for this fitting. This leads to values of $\alpha{_{i}}$ that better reproduce the saturation effect of a large collection of various lines of a given ion.

\subsection{Bound-free transitions}
\label{sec:bf}

The equations for ${g}_{\rm rad}$ due to bound-free transitions are very different in nature to those described above. First, the momentum acquired through photoionisation is gained by the newly created ion. Also, part of this momentum is transferred to the ejected electron \citep[see for instance][]{MassacrierMaEl1996r}. As in most of the methods presented in Sec.~\ref{sec:intro}, this last effect is neglected in the SVP method. Moreover and in order to develop algebraic equations, the ionisation cross-sections are approximated by a power law similar to the one found for hydrogenic ions (i.e. proportional to $\nu^{-3}$). 

Following the paper by \citet{AlecianAl1994}, \citet{AlecianAlLe2002} and \citet{LeBlancLeAl2004} developed the SVP equations for bound-free transitions (see equations 4, 5, 6 and 9 of \citealt{LeBlancLeAl2004}), the acceleration due to bound-free transitions of ion $i-1$ that applies to ion $i$ is approximated by:

\begin{equation}
g_{\rm{i,cont}}\approx{a_i\left[{g_{\rm{i,cont}}}\right]}_{\rm{Eq.4(2004)}}{\left({{{\chi}
\over{1+\chi}}}\right)}^{b_{i}},
\label{gcontchi}
\end{equation}

\noindent where $\chi$ is the abundance of the element relative to its solar
value in number, and $\left[{g_{\rm{i,cont}}}\right]_{\rm{Eq.4(2004)}}$ refers to the equation (4) of \citealt{LeBlancLeAl2004} (from which we have removed the term $a_i$ and put it at the beginning of the right hand term). In Eq.~\ref{gcontchi} the two parameters $a_i$ and $b_i$ have to be determined by a fitting procedure (at the same time when $\alpha{_{i}}$ is obtained, see Sec.~\ref{sec:fit}). By default these parameters are respectively set to 1 and 0~.

It is important to state that Eq.~\ref{gcontchi} is much more inaccurate as compared to Eq.~\ref{gline}. In an ideal situation, one should estimate the $g_{\rm{i,cont}}$ from the detailed photoionisation cross-sections. But they are not exhaustively available, and the physics of photoionisation is much more complex than for bound-bound transitions.
However, it is very important that $g_{\rm{i,cont}}$ not be neglected, because in case of elements extremely saturated like CNO and even Fe, $g_{\rm{i,cont}}$ prevents the radiative acceleration becoming extremely small. This is why we are convinced that it is better to have an inaccurate estimate of $g_{\rm{i,cont}}$, than simply neglecting it. Moreover, $g_{\rm{i,cont}}$ is most often much smaller than $g_{\rm{i,line}}$ for typical stellar chemical abundances, and also much smaller than gravity. Therefore, the consequences of its inaccuracy are generally marginal. 

The inaccuracy of $g_{\rm{i,cont}}$ discussed above is partly corrected through our fitting procedure for $a_i$. The second parameter $b_i$ was added to take into account possible saturation effects for bound-free transitions. Its value is also obtained through the fitting procedure described in Sec.~\ref{sec:fit}.

\subsection{Total accelerations}
\label{sec:total}

The diffusion of particles for an element depends on the total radiative acceleration. Since the SVP method gives radiative accelerations of ions, they have to be combined to obtain the total acceleration of the element. This expression for the combined effect of all ions is given by equation (7) of \cite{LeBlancLeAl2004}:

\begin{equation}
g_{\rm tot}={{\sum\limits_{i}{w_{i}N_{i}\left({g_{\rm i,line}+g_{\rm i,cont}}\right)}}\over{
\sum\limits_{i}{w_{i}N_{i}}}},
\label{gtot}
\end{equation}

\noindent
where $N_{i}$ is the density of ion $i$ in number, and  $w_i$ a weight, which in some cases may be set equal to the ion diffusion coefficient, or just equal to unity (the default value in OPCD codes). In this work, the weights $w_i$ are generally set equal to $1$, however we have found that the results are improved with $w_i=1.5$ for ions in noble gas configurations (this is the option we adopted in the results we present later). For $N_{i}$, we have simply computed them with the Saha-Boltzmann equations using a simplified list of the ions' energy levels. The calculation of ions' populations are often missing in evolution numerical codes. Therefore, the SVP tables are provided with the data and codes needed for their calculation (see Sec.~\ref{sec:data}).

\section{Recent improvements of the SVP method}
\label{sec:improv}

As recalled in Sec.~\ref{sec:svp}, there are six parameters per ion in SVP tables (those that are used by the end user). It should be noted that the three parameters $\phi{_{i}}^{*}$, $\psi{_{i}}^{*}$ and $\xi{_{i}}^{*}$ are directly obtained through atomic data for each ion, and depend on the plasma conditions, but not on the abundance of the ion under consideration. The values of these three parameters are not modified by the fitting procedure described in Sec.~\ref{sec:fit}, and are therefore provided as is in the SVP tables. The fitting procedure determines the best values of the other three parameters ($\alpha{_{i}}$, $a_{i}$ and $b_{i}$).

Several improvements have been brought to the calculations of the six SVP parameters since the results of \cite{LeBlancLeAl2004}. They are described in this section.

\subsection{Calculation of the three unfitted parameters}
\label{sec:unfit}

In the results presented here, we have significantly improved the way opacities are interpolated from those found in the OP tables, and also, the way they are included in the calculation of these parameters. In the previous version of the tables, opacities were first interpolated from OP tables (in a $T$, $N_e$ grid) for each stellar layer of the numerical model before being used in the formulae of the $\phi{_{i}}$, $\psi{_{i}}$ and $\xi{_{i}}$ parameters\footnote{We recall that $\phi{_{i}}^{*}$, $\psi{_{i}}^{*}$ and $\xi{_{i}}^{*}$ are respectively equal to $\phi{_{i}}$, $\psi{_{i}}$ and $\xi{_{i}}$ obtained for the layer where the ion $i$ is near to its maximum relative population. Therefore, there is a unique set of $\phi{_{i}}^{*}$, $\psi{_{i}}^{*}$ and $\xi{_{i}}^{*}$ parameters for a given stellar model, while there are as many sets of $\phi{_{i}}$, $\psi{_{i}}$ and $\xi{_{i}}$ as there are layers in the model.} (see Eq. 1 to 15 of \citealt{AlecianAlLe2002}). In the present version, we first compute four sets of  $\phi{_{i}}$, $\psi{_{i}}$ and $\xi{_{i}}$ on the mesh points of the OP tables for a set of $T$ and $N_e$ surrounding the point corresponding to the layer of the stellar model. In a second step, we interpolate from these parameters to get the three final parameters at the temperature and density of the layer. The resulting improvement is that the values of the $\phi{_{i}}$, $\psi{_{i}}$ and $\xi{_{i}}$ parameters vary in a much smoother way from layer to layer than in the previous version, and therefore, these more accurate values are used to determine $\phi{_{i}}^{*}$, $\psi{_{i}}^{*}$ and $\xi{_{i}}^{*}$ for the layer where the ion $i$ is near its maximum relative population.

Another improvement in the new version of the SVP method is that we now consider effective quantum numbers of bound-bound transitions, rather than main quantum numbers. This is especially important to estimate the collisional impact on the line widths for the Lorentz profile (term $\gamma_{il}$ in Eq. 7 of \citealt{AlecianAlLe2002}, see also Eq.9 of \citealt{AlecianAlAr1990a}). This change improves the parameter $\psi{_{i}}^{*}$ that accounts for saturation effects of atomic lines.

We have also reconsidered the estimate of $\xi{_{i}}^{*}$. Indeed, this parameter, which is obtained through a first order Taylor expansion (see Eq. 13 and 15 of \citealt{AlecianAlLe2002}) may introduce some error as discussed in the Sec.~3.2 of \citet{AlecianAlLe2002}. In the light of the results of many numerical tests, we decided to restrict more tightly its domain of validity, and set it equal to zero outside this domain. This is why this parameter is often equal to zero in the new SVP tables.
 
\subsection{Fitting of the three other parameters}
\label{sec:fit}

The fitting procedure for the three parameters $\alpha{_{i}}$, $a_{i}$ and $b_{i}$ has also been improved. These parameters are now found by a simultaneous fit of the SVP accelerations to those of \cite{SeatonSe2007u} for five abundances (from -2 dex to +2 dex relative to a reference abundance, which is here the solar abundance). The errors related to each abundance are now weighted by a factor of 5 for the reference abundance, a factor of 3 for $\pm$ 1 dex and a factor of 1 for $\pm$ 2 dex relative to the reference abundance to lead to an overall error that is minimised through our fitting procedure. Remember that at the end there is a unique set of 6 parameters per ion whatever the value of the abundance of the considered element (provided that it remains in the validity domain of $\pm$ 2 dex). Since the relative weights for the errors at extreme abundances are smaller, the uncertainty of our results increases as the abundance differs from its reference value, especially for abundances nearing  $\pm$ 2 dex relative to reference abundances. However, this leads to better results near solar values.

In order, in some instances, to better fit the ${g}_{\rm rad}$ obtained by the SVP method to those of \cite{SeatonSe2007u}, the authorised domain for the fitted parameters in our new fitting procedure was expanded. The fitting procedure was therefore applied to the following domains: -1 $ \leq \alpha{_{i}} \leq$ 0, 0 $ \leq a{_{i}} \leq$ 2 and -2 $ \leq b{_{i}} \leq$ 0 .

\section{Sample of results obtained with the new SVP tables}
\label{sec:sampl}

\begin{figure*}
\includegraphics[width=16cm]{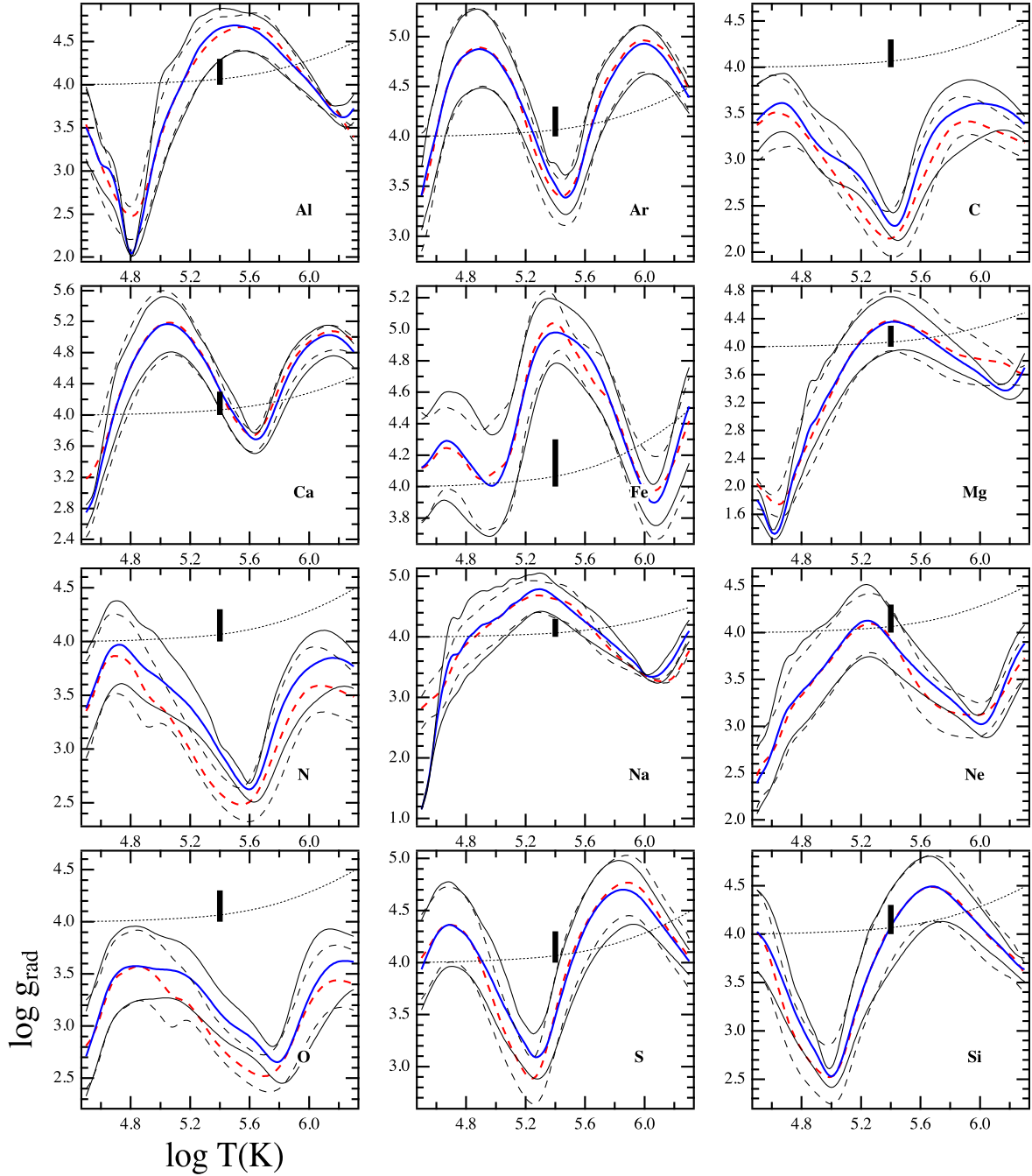}
\caption{
Radiative accelerations for the 2.0~M$_{\odot}$ main-sequence model ($T_{\rm eff}\approx 8210$K, $\log{g}=4.0$). The logarithm of accelerations are plotted vs. the logarithm of the temperature of the layers for 12 metals (the name of each element is indicated at the right bottom corner of each plot). Notice that left axis scale is different for each element. Solid curves show accelerations given by the OPCD codes, the dashed curves are those given by the SVP method. The blue and red curves are for solar abundance, the two other sets of curves are for 1/10 (upper ones) and 10 times solar abundance. The dotted curve is the local gravity as given by the model. The thick vertical bar (positioned at $\log T \ $= 5.4) shows 0.3 dex on the left axis which is the maximum desired error for ${g}_{\rm rad}$ (see text).}
\label{fig:8210}
\end{figure*}

\begin{figure*}
\includegraphics[width=16cm]{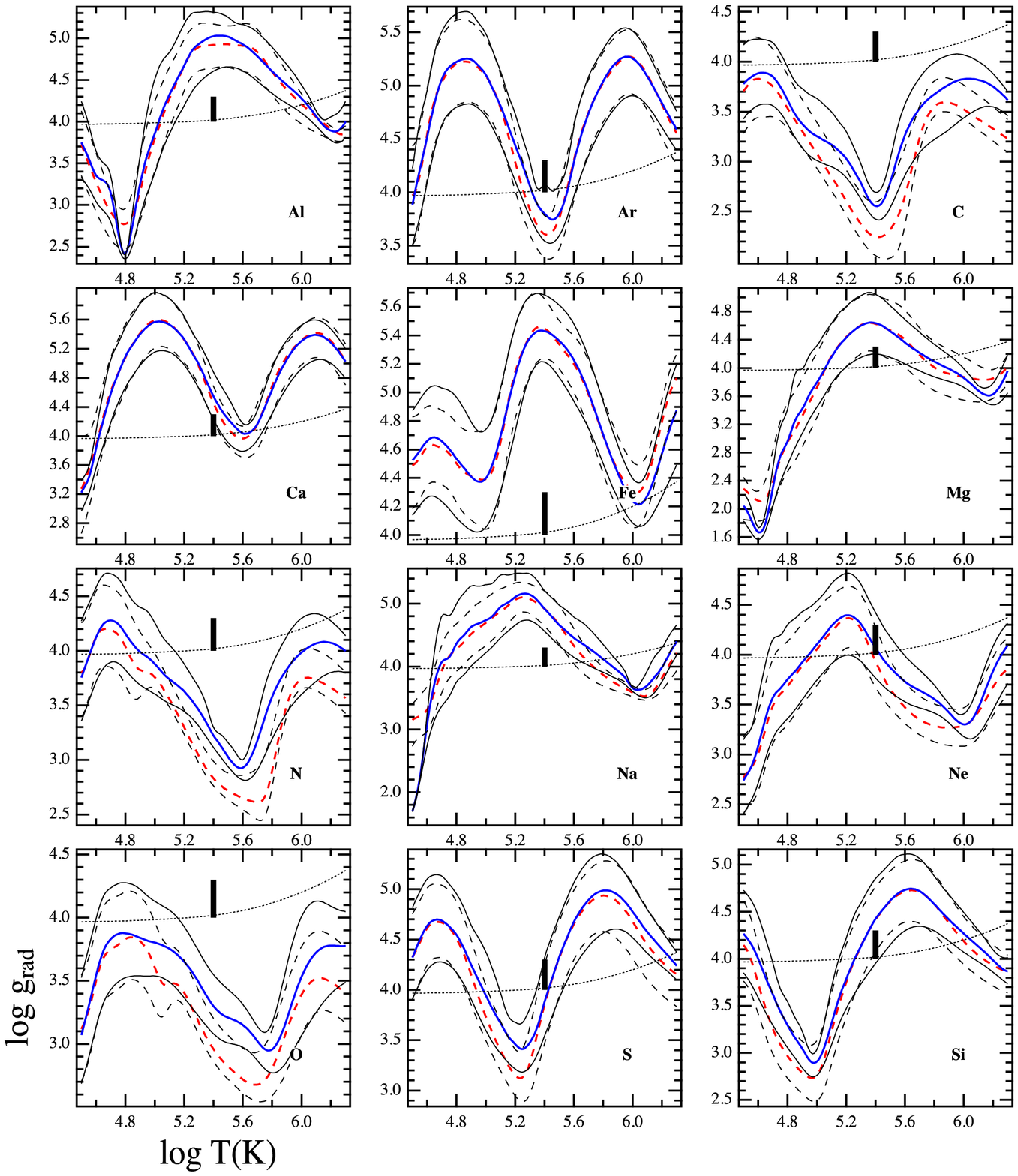}
\caption{
Same as Fig.~\ref{fig:8210} for the 3.5~M$_{\odot}$ main-sequence model ($T_{\rm eff}\approx 12240$K, $\log{g}=4.0$).}
\label{fig:12240}
\end{figure*}

The SVP parameters discussed in this work have been obtained using stellar models computed with the evolution code CESTAM \citep{MarquesMaGoLeetal2013r,DealDeAlLeetal2018} (kindly communicated by M. Deal) with stellar masses from 1 to 10 solar masses and being in about the middle of their life on the main sequence. These models have been computed using the solar abundances of \citet{AsplundAsGrSaetal2009}. For the sake of conciseness, we show in Fig.~\ref{fig:8210} and \ref{fig:12240} the results for 2.0~M$_{\odot}$ ($T_{\rm eff}\approx 8210$K, $\log{g}=4.0$) and 3.5~M$_{\odot}$ ($T_{\rm eff}\approx 12240$K, $\log{g}=4.0$) models, but results for all the models are shown in the website presented in Sec.~\ref{sec:data}. We have considered the 12 elements (C, N, O, Ne, Na, Mg, Al, Si, S, Ar, Ca and Fe) for which the Opacity Project has provided detailed atomic data and radiative accelerations. We have not considered Ni for which the OPCD radiative accelerations are based on extrapolations from iron.

For both models presented here, accelerations obtained with the SVP method are relatively close to those obtained using OPCD codes, and the deviations are generally smaller than the acceptable error of 0.3 dex (see the vertical bars in the figures), except for some accelerations much smaller than gravity. In that last case, large errors for ${g}_{\rm rad}$ have no significant consequences on the diffusion process, since in this precise case the dominant force is gravity. This is especially true for CNO that have large solar abundances (very saturated lines) and small atomic numbers. The latter one is not a very favourable property regarding one of the approximations used for SVP, i.e. to consider the value of parameters at layers where populations of ions are close to their maximum. This is because light elements have few ionisation states, layers for which ions have their maximum population are more distant from each other, and so, there are a smaller number of SVP parameters for the element than for heavier ones. Also, the assumption that consists in considering elements as trace elements with regards to the background opacity is possibly not well respected for elements like CNO or Fe, especially when they are overabundant.

For most metals and models, the SVP method gives a satisfactory level of accuracy for computing radiative accelerations. Our results are particularly precise for elements with small solar abundances (i.e. Ar or Si for example), especially where ${g}_{\rm rad}$ is larger than local gravity. As discussed above, results for elements with large solar abundances such as CNO give less accurate results. Another factor that intervenes for such elements is that due to the very high saturation of their atomic lines, contribution of bound-free transitions to ${g}_{\rm rad}$ may dominate. Even if this has no serious consequence, as explained previously, it is an extra reason to question the accuracy of our computed ${g}_{\rm rad}$ for CNO, since, according to the discussion in Sec.~\ref{sec:bf} momentum transferred through photoionisation to atoms is not accurately estimated in the SVP method. One may also notice a dip, which does not appear in the OPCD calculation, for radiative acceleration of Fe at $\log T \approx 5.6$ and low abundances that is not well explained. We suspect that this may be due to atomic data, which were downloaded from an old version of OP data. This will be checked when the new release of the OP atomic data will be available. They are in preparation (private communication of F. Delahaye), and will be corrected in a future release of SVP tables. Notice, however, that the dip is smaller than 0.3 dex.

\begin{figure}
\includegraphics[width=8cm]{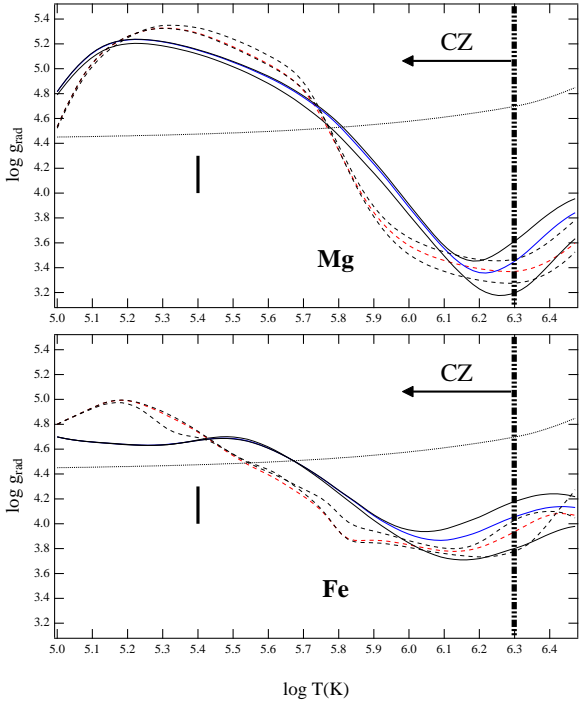}
\caption{A sample of SVP radiative accelerations for the 1.0~M$_{\odot}$ model, which corresponds to the lower limit in stellar mass of the SVP tables provided. The curves have the same meaning as in Fig.~\ref{fig:8210} and \ref{fig:12240}, except for the vertical line at $\log T \approx 6.3$ that marks the bottom of the superficial convection zone.}
\label{fig:Layout1Msun}
\end{figure}

For the lowest mass model considered (1.0~M$_{\odot}$), the SVP method doesn't work as well as for models for larger masses.  
Figure~\ref{fig:Layout1Msun} shows the best (Mg) and the worst (Fe) cases to illustrate this observation. Nevertheless, for a 1.0~M$_{\odot}$ star having the age of the Sun, all the superficial layers above  $\log T \approx 6.3$ are convective and therefore, atomic diffusion is completely inefficient in these layers, and an error on ${g}_{\rm rad}$ has no consequence. We have decided to provide on our website the table of parameters even for this model to have a more complete interval of masses (1-10~M$_{\odot}$). Notice however, that deeper than  $\log T \approx 6.3$, where diffusion can be efficient (even if it is with very small timescale), the radiative acceleration provided by our SVP table are not far from the 0.3 dex of accuracy we require. This may be important, since as computed by \citet{TurcotteTuRiMietal1998}, even if ${g}_{\rm rad}$ for Fe is small below the convection zone (about $1/5$ of $g$) it may significantly affect the superficial abundance of Fe.

\section{Data and codes for numerical applications}
\label{sec:data}

The main outcome of the present study is the set of new SVP tables providing the parameters $\varphi{_{i}}^{*}$, $\psi{_{i}}^{*}$, $\xi{_{i}}^{*}$, $\alpha{_{i}}$, $a_{i}$ and $b_{i}$ that are discussed in Sec.~\ref{sec:svp}. As seen in Sec.~\ref{sec:svp}, these parameters depend on the stellar model, however this dependency is not very strong. Therefore, seventeen stellar models well distributed in mass, are enough to cover the 1-10~M$_{\odot}$ domain. This leads to 17 SVP tables from which the parameters will be interpolated for any mass being inside that domain (see Sec.~\ref{sec:tables}).

Although the SVP parameters enter a few simple formulae (Eq.~(\ref{gline}) to (\ref{gcontchi})) without particular difficulties in implementing the method in numerical codes, it may represent a significant programming effort. To help the end user in applying the SVP method, we now provide a standalone program which shows how to use these SVP tables. All the source codes (written in Fortran 90) and data may be downloaded from the website {\it http://gradsvp.obspm.fr} in the form of the svp\_standalone.tar.gz file that contains the following files and subdirectories:
\begin{description}
\item --~~\verb'READme.txt' contains detailed description of the content, and instructions of use;
\item --~~\verb'srcp/' contains source codes written in Fortran 90 (see Sec.~\ref{sec:democ});
\item --~~\verb'datai_SVP_v1/' contains data needed to compute radiative accelerations (see Sec.~\ref{sec:tables});
\item --~~\verb'data_standalone/' contains data used by the demo program svp\_standalone.f;
\item --~~\verb'fgrp' is a C shell compilation script (gfortran).
\end{description}.

\subsection{The standalone demo code and the SVP package}
\label{sec:democ}
The standalone program consists of two entities, the standalone demo code (file svp\_standalone.f), which contains the main program and its specific subroutines needed to run the demo, and a set of 3 files (svp\_codes.f, mod\_svp.f, mod\_donvar.f) that constitutes the SVP package that also has to be used in a stellar evolution code following the way it is used by the demo program.

After compilation, the demo standalone program is ready for use. At the launch, the program asks to choose a stellar mass among a list of available values that corresponds to the content of the library of limited static models computed with the CESTAM evolution code as previously mentioned and used in Sec.~\ref{sec:sampl} (from 1 to 10 solar mass; one needs to enter the value of the mass $\times 100$). The essential output of the standalone program is a file containing radiative accelerations for all layers and elements for which the CESTAM models considers diffusion. When the SVP package is implemented in another stellar evolution code, the calling initialisation routine must provide to the package\footnote{Using fortran MODULEs.} the mass of the considered star (any mass between 1 and 10 solar mass), and a list of elements (or isotopes). The package will determine which ones of theses elements have available SVP parameters in the SVP tables, and will compute the radiative accelerations for these elements. The subroutine computing radiative accelerations must be called for each layer and each time step. Note that a default setting -- an option that may be adjusted by the user in the source code (in the file svp\_code.f) -- is that radiative acceleration of an element without available SVP parameters is set equal to local gravity. It means that radiative acceleration and gravity cancel each other for this element, preventing it from diffusing due to an external force. Indeed, it is generally better to prevent an element from diffusing when radiative acceleration is unknown than to let gravity act alone. However, this could be an undesired option for some elements, for instance for helium that is not present in SVP tables. Actually, even if in some cases the radiative acceleration of HeII may be non-negligible, it may be justified in most cases to assume that radiative acceleration on He is negligible and and let it to diffuse due to gravity alone\footnote{HeI being in noble gas configuration, radiative acceleration is smaller than gravity.}. In the present version of the package, it is up to the user to force the output acceleration for some elements to be different from the standard output of the SVP package.  
Notice that, to be worth computing radiative accelerations in an evolution code, it is mandatory that this code is able to compute the detailed abundance evolution and opacities at each time step. Evolution codes that use an average metal abundance to estimate an average opacity cannot account with the effect of  abundance changes on radiative accelerations, which necessarily leads to meaningless abundance stratification.

\subsection{The SVP tables and preprocessed atomic data}
\label{sec:tables}
The SVP parameters are provided for a collection of stellar masses inside the \verb'datai_SVP_v1/SVP_tables' subdirectory. There are 17 files used by the SVP package to determine through interpolation the best suited parameters for the stellar mass considered by the calling program\footnote{The package does not need to interpolate when invoked by the standalone demo code, since the tables have been built for the very models than those being inside the data\_standalone library.}. In the present version of the data, SVP parameters are provided for the 12 elements (C, N, O, Ne, Na, Mg, Al, Si, S, Ar, Ca and Fe) for which the Opacity Project has provided detailed atomic data. In a future release of SVP tables, we will add scandium \citep{AlecianAlLeMa2013}. Other future releases will be provided according to the future releases of OP data. The SVP package allows easy updates of the input data. 

Since radiative accelerations of the various isotopes of a given element are almost identical \footnote{Radiative acceleration depends strongly on the saturation effect of spectral lines (related to abundances), which determines the available photons that can transfer their momentum to the element. Because, the radiation frequency of atomic transitions of isotopes are almost identical, all isotopes {\it see} the same flux of photons, and so, they have the same radiative acceleration. In the SVP package, the mass difference between isotopes is neglected.}, tables for the 12 elements apply also to their isotopes.

The \verb'datai_SVP_v1/' subdirectory contains a subdirectory: \verb'fused_levels/' that contains 173 files (one per ion) with the atomic data needed to compute ions relative populations. Because stellar evolution codes generally do not compute relative population of ions, the SVP package has to be self-sufficient for that need. These small files contain atomic data for the energy levels of ions. Since when calculating ions relative population, the list of energy levels does not need to be known in a very detailed way, and to reduce the amount of data to handle\footnote{In recent atomic databases, the number of provided energy levels for each ion is generally huge.}, these data have been preprocessed. That process consists in fusing energy levels together to drastically reduce their number without significant loss of accuracy for the calculation of ions population (and partition functions). Levels with an energy above 2/3 of the ionisation potential are not considered, because these levels are not significantly populated in stellar layers where the considered ion may contribute to the total radiative acceleration [Eq.(\ref{gtot})].

\section{Additional comments}
\label{sec:comments}

\subsection{Abundances}
\label{sec:abund}

The results presented in Fig.~\ref{fig:8210} to \ref{fig:Layout1Msun} have been obtained for models computed with solar abundances \citep{AsplundAsGrSaetal2009}, while the presented ${g}_{\rm rad}$ have been calculated for several abundances for each element (by a factor of 0.1, 1., 10. relative to solar in all the layers, one metal at a time). Actually, to compute these ${g}_{\rm rad}$ we consider metals as trace elements concerning the stellar structure and so, assume that the models are not affected by the these abundance changes. However, the value of ${g}_{\rm rad}$ takes the abundance changes into account through  $C_{i}$ in Eq.~\ref{gline}. These results may also be obtained using the demo code. Of course, when the SVP package is implemented in an evolution code (see Sec.~\ref{sec:democ}), at each time step the ${g}_{\rm rad}$ calculation is consistent with local abundances calculated by the evolution code.

\subsection{Validity domain}
\label{sec:valid}
Concerning the validity domain, our SVP parametric method has been tested for POP I main-sequence stars. We have found that the method works quite well for stellar masses larger than 1~M$_{\odot}$ and we have tried to verify the accuracy of the method for lower masses (for 0.8 and 0.9~M$_{\odot}$), but the obtained radiative accelerations are too far from those obtained by the OPCD codes to be satisfactory. This appears to be linked to the fact that, for most elements, layers where ions have their maximum relative populations, are too distant from each other.  The SVP method has not been checked for masses larger than 10~M$_{\odot}$, and for other models than for POP I main-sequence stars, however we cannot exclude that it may work for other stellar types as for instance horizontal-branch stars that may have internal structure more or less similar to main-sequence stars at the same effective temperature. But, this will certainly require to build specific SVP tables. Also, SVP tables are not computed for the neutral stages of metals, and not for H and He. Finally, the SVP method can only be applied in optically thick regions.

\subsection{Sensitivity to models}
\label{sec:models}
One of the main foundations of our method is that the SVP parameters ($\varphi{_{i}}^{*}$, $\psi{_{i}}^{*}$ and $\xi{_{i}}^{*}$) presented in details in Sec.~\ref{sec:svp} are only faintly dependent on the plasma conditions and on local abundances. A consequence of this property is that these parameters are not very sensitive to local changes of temperature, density and abundances. Therefore, the same SVP table may be used to compute radiative accelerations for the star during its evolution on the main-sequence. This is also true when models are computed by various evolution programs, as far as the produced models do not present large differences compared to standard models, which is generally the case. This is why the tables provided here and computed using the CESTAM code can be used by other evolution codes. For stars with very different metallicity (POP II stars for instance), or peculiar structure, it will be preferable to build specific SVP tables. Such extensions of tables will be provided on our website {\it http://gradsvp.obspm.fr} in the future.

\subsection{Opacities}
\label{sec:opac}
For opacities that are used to prepare the SVP tables (see \citealt{AlecianAlLe2002}, and Sec.~\ref{sec:fit} for the fitting), the most recent available version of The Opacity Project data (TOPbase version OPCD3.3, \citealt{SeatonSe2005}), where the inner-shell configurations \citep{Badnelletal2005} are included in the monochromatic opacities and ${g}_{\rm rad}$ calculations of \cite{SeatonSe2005} and \cite{SeatonSe2007u}, has been used here. However, detailed atomic line data used to compute $\phi{_{i}}$, $\psi{_{i}}$ and $\xi{_{i}}$ parameters are still those provided by the old version of OP \citep{SeatonSeZeTuetal1992}.

\subsection{Compatibilty with the OPCD codes}
\label{sec:opac}
Concerning the implementation of the SVP package in evolution codes, as we point out in Sec.~\ref{sec:intro}, the SVP method brings the important advantage of high speed or possibility of extension to other elements than those considered in opacity databases, and should have enough accuracy in most cases. However, the method has some limitations such as the stellar mass domain (presently from 1 to 10~M$_{\odot}$), or stellar types (presently main-sequence). Therefore, we suggest that the SVP package could be implemented in parallel with the OPCD codes, so it could be possible to choose either of these methods according to the needs. One could even use both (according to the element) in a mix mode, since both methods are compatible and consistent. However, we have never tested such an option.

\section{Summary}
\label{sec:conc}
In this work we discuss, in the light of several years of use, our SVP parametric method for fast numerical calculation of radiative accelerations in the interior of main-sequence stars, and we present its current status. We also provide new improved tables of the SVP parameters for a relatively large stellar mass interval (from 1 to 10~M$_{\odot}$). The improvements brought to the SVP method are described in Sec.~\ref{sec:improv}. We now also provide a standalone program, which is a demo program for the use of the SVP tables. It comes with all the necessary codes in view of the implementation of the SVP method in existing numerical stellar evolutionary programs. All these data and codes are available and freely downloadable from the website {\it http://gradsvp.obspm.fr}, which is described in Sec.~\ref{sec:data}. A special effort has been done to make these data and codes easy to use.

The main outcome of this work is that the SVP method is efficient in terms of rapidity (about 1000 times faster than the OPCD method), flexibility and accuracy to be worth implementing in existing stellar evolutionary codes. We notice that it may be included in a program in which the codes provided by OPCD are already implemented. In the near-future, we foresee adding SVP parameters for other elements (starting with Sc and Ni). This will be done through updates of our website {\it http://gradsvp.obspm.fr}.

\section*{Acknowledgements}

We would like to thank M. Deal for having provided the stellar models used in this paper, and that are available in the numerical package of data and codes of our website {\it http://gradsvp.obspm.fr}.
This work was supported by a grant from the Natural Sciences and Engineering Research Council of Canada.
Part of the calculations were done on the supercomputers {\it briarree and graham}, under the guidance of Calcul Qu\'{e}bec and Calcul Canada. The use of these supercomputers is funded by the Canadian Foundation for Innovation (CFI), NanoQu\'{e}bec, RMGA and Research Fund of Qu\'{e}bec - Nature and Technology (FRQNT). We acknowledge the financial support from the Programme National de Physique
Stellaire (PNPS) of CNRS/INSU, France.

%%%%%%%%%%%%%%%%%%%%%%%%%%%%%%%%%%%%%%%%%%%%%%%%%%
\section*{Data Availability}

The data underlying this article are available at http://gradsvp.obspm.fr/datasheet\_2020.html, and can be freely accessed.

%%%%%%%%%%%%%%%%%%%%%%%%%%%%%%%%%%%%%%%%%%%%%%%%%%



\bibliographystyle{mnras}
\bibliography{new_svp}

%\cite{*}

%%%%%%%%%%%%%%%%%%%%%%%%%%%%%%%%%%%%%%%%%%%%%%%%%%

% Don't change these lines
\bsp	% typesetting comment
\label{lastpage}
\end{document}